\documentclass{article}
\usepackage{graphics}
\usepackage{epsfig}
\usepackage{amsmath}
\usepackage{graphicx}

\makeatletter
\newcommand{\row}[1]%
{\mathord{\buildrel{\lower3pt%
\hbox{$\scriptscriptstyle\rightarrow$}}\over #1}}

\newcommand{\dyadic}[1]{\mathord{\dyadic@rrow{#1}}}
\newcommand{\dyadic@rrow}[1]{
\begin{picture}(12,12)(-1,0)
\put(-3,12){\makebox(0,0)[t]{$\scriptscriptstyle\downarrow$}}
\put(-3,13){\makebox(0,0)[l]{$\scriptscriptstyle\longrightarrow$}}
\put(5,0){\makebox(0,0)[b]{$#1$}}
\end{picture}
}

\newcommand{\bra}[1]{\bigl\langle #1 \bigr|}
\newcommand{\ket}[1]{\bigl| #1 \bigr\rangle}

\begin{document}

\begin{center}
\bigskip {\Large  Dynamics of multi-modes  maximum entangled coherent state over
 amplitude damping channel}

\bigskip A. El Allati $^{a,b}$, Y. Hassouni $^{a}$
 and  N. Metwally $^{c}$\\
 $^{a}$ Facult\'e des
Sciences, Laboratoire de Physique Th\'{e}orique URAC 13,
Universit\'e Mohammed V - Agdal.
Av. Ibn Battouta, B.P. 1014, Rabat, Morocco\\[0pt]
$^{b}$ The Abdus Salam International Centre for Theoretical Physics,
Trieste, Italy\\
$^{c}$Mathematics Department, College of Science,  University of
Bahrain, P.O. Box, 32038 Bahrain. \\
$^{c}$Mathematics Department, Faculty of Science,  South Valley
University, Aswan, Egypt.
\end{center}

\vspace{0.5cm}

\begin{abstract}
The dynamics of maximum entangled coherent state travels through
an amplitude damping channel is investigated. For small values of
the transmissivity rate the travelling state is very fragile to
this noise channel, where it suffers from the phase flip error
with high probability. The entanglement decays smoothly for larger
values of the transmissivity rate and speedily for smaller values
of this rate. As the number of modes increases, the travelling
state over this noise channel loses its entanglement hastily. The
odd and even states vanish at the same value of the field
intensity.

\bigskip

\end{abstract}

\vspace{0.5cm}

\section{Introduction}
 Entanglement is one of the fundamental
properties of quantum information theory, where it has been
considered as a nonclassical resource for many applications as
quantum teleportation \cite{Ben} and super dense coding
\cite{Ben1}. To achieve these tasks with high efficiency one needs
maximum entangled states and perfect local operations, which are
very difficult to be established in the real word. Therefore,
investigating the dynamics of entanglement in the presence of
imperfect circumstance is very important in the context of quantum
information processing. For example, the dynamics of multiparities
entanglement under the influence of decoherence is investigated in
\cite{Carvahlo,Konrad}. The dynamics of entangled atoms interact
with a deformed cavity mode is investigated by Metwally
\cite{Metwally1}.

Coherent states  play  important roles in  many fields of physics,
specially in quantum technologies and quantum optics \cite{Deu}.
For example, two entangled coherent states are used to realize an
effective quantum computation\cite{Jen} and quantum teleportation
\cite{Enk}.  Allati and et al \cite{allati1} have suggested a
system of   three modes coherent state and used it to perform
quantum teleportation. Communication via entangled coherent
quantum network is investigated in \cite{allati2}, where it is
shown that the probability of performing successful teleportation
through this network depends on its size.

Entanglement properties of an optical coherent entangled state
consists of two entangled modes under amplitude damping channel is
discussed  by Wickrt\cite{Ricardo}. The dynamics of the GHZ state
through the amplitude damping channel is investigated by  Konrad
et. al \cite{Konrad}. This motivates us to investigate the
entanglement properties of a class of  maximum entangled coherent
states consist of three modes pass through a damping channel.
Also, we study the dynamics of a multi-entangled coherent state
passes through this noise channel. The effect of this channel
equivalence to a photon absorption  followed by a phase flip
operator. The suppressing of the travelling state  over this
channel is discussed, where we quantified  the bound entanglement
of the output state as well as the survival amount of
entanglement.

The paper is organized as follows: In Sec.2, we review the
suggested entangled  muti-modes coherent state, MMCS  the amount
of entanglement  over a perfect environment is quantified
\cite{allati1}. The entanglement of the MMCS over an amplitude
damping channel is investigated in Sec.3, where we quantify the
bound of entanglement for a maximum entangled state consists of
three modes.  The dynamics of  multi-modes entangled coherent
state passes through the damping channel is discussed. Finally, we
summarize our results in Sec.4.

\section{Perfect environment}

Entangled coherent states have been proposed as an important
 resource in quantum information  processing, ensuring
or teleporting an unknown quantum states. These sates can be
written as function of the Fock state \cite{Gil} as,
\begin{equation}
\bigl| \pm\alpha \bigr\rangle=\exp(-2|\alpha|^2)\sum_{n=0}^{\infty}{\frac{%
(\pm\alpha)^n}{\sqrt{n!}}\bigl| n \bigr\rangle}.
\end{equation}
The coherent state can be generated from the vacuum state
$|0\rangle$, by the displacement operator $D(\alpha)=exp(\alpha
\hat{a}^{\dag}-\alpha^{*}\hat{a})$, where $\hat{a}^{\dag}$ and
$\hat{a}$ are bosons creation and annihilation operators
respectively. Among of the  properties  of these states is the
non-orthogonality, and the overlap of two coherent states
$|\pm\alpha\rangle$ is $\langle\alpha|-\alpha\rangle =
exp(-2|\alpha|^{2})$ which becomes orthogonal by increasing the
amplitude $|\alpha|$. Two coherent states can be used as basis
states of a logical qubit where $|0\rangle_{L}=|\alpha\rangle$ and
$|1\rangle_{L}=|-\alpha\rangle$. One form of the entangled
coherent states between three modes can be written as,
\begin{figure}[b]
\begin{center}
  \includegraphics[scale=0.4]{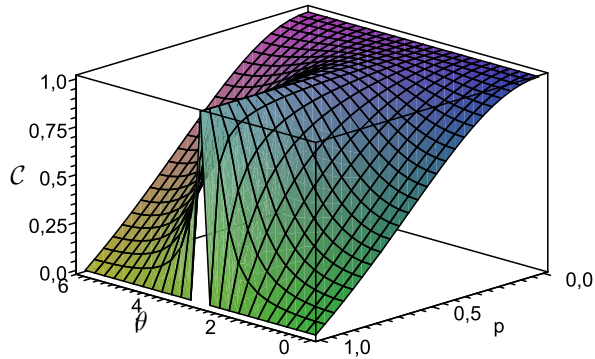}\put(-178,25){$\theta$}
  \put(-225,80){$\mathcal{C}$}
  \caption{Concurrence for the coherent states for a range
  of $\theta$ and $p$, with $p=\langle\alpha|-\alpha\rangle^{2}$.}\label{conc1}
\end{center}
\end{figure}

\begin{eqnarray}  \label{channel}
\rho_{\alpha} &=&\frac{1}{N_{\theta
}^{2}}\Bigl\{\bigl|\sqrt{2}\alpha ,\alpha ,\alpha
\bigr\rangle_{345}\bigl\langle\sqrt{2}\alpha ,\alpha ,\alpha
\bigr|+e^{-i\theta} \bigl|\sqrt{2}\alpha ,\alpha ,\alpha
\bigr\rangle_{345}\bigl\langle-\sqrt{2}\alpha ,-\alpha ,-\alpha
\bigr|  \nonumber \\
&+&e^{i\theta}\bigl|-\sqrt{2}\alpha ,-\alpha ,-\alpha
\bigr\rangle_{345}
\bigl\langle\sqrt{2}\alpha ,\alpha ,\alpha \bigr|  \nonumber \\
&+&\bigl|-\sqrt{2}\alpha ,-\alpha ,-\alpha \bigr\rangle_{345}
\bigl\langle-\sqrt{2}\alpha ,-\alpha ,-\alpha \bigr|\Bigr\},
\end{eqnarray}
where $N_{\theta }=\sqrt{2(1+ e^{-16|\alpha |^{2}}cos(\theta))}$
is the normalization factor. If we  set  $\theta=\pi$ in
(\ref{channel}), one obtains a maximum entangled state defined by,
\begin{equation}\label{max}
\rho^{-}_{\alpha}=\ket{\psi^{-}_{\alpha}}\bra{\psi^{-}_{\alpha}},\quad
\ket{\psi^{-}_{\alpha}}=
\frac{1}{\sqrt{N_{\alpha}}}(\ket{\sqrt{2}\alpha,\alpha,\alpha}-\ket{-\sqrt{2}\alpha,-\alpha,-\alpha})
\end{equation}
where  $N_{\alpha}=2(1-e^{-8|\alpha|^{2}})$ is the normalization
factor. We use the  concurrence to quantify entanglement between
two qubits, which is denoted by $\mathcal{C(\ket{\psi_{\alpha}}})$
as \cite{Dur},
\begin{equation}
\mathcal{C}^{1/23}(\ket{\psi_{\alpha}})=\frac{1-exp(-8|\alpha|^{2})}{1+exp(-8|\alpha|^{2})cos(\theta)}.
\end{equation}
 Fig.(\ref{conc1}), describes the dynamics of entanglement contained
in  the state $\ket{\psi_{\alpha}}$ as function  of $\theta$ and
$|\alpha|$. It is clear that, at $\theta=\pi$ the concurrence
$\mathcal{C}=1$ namely the   entanglement is maximum and is
independent of $|\alpha|$. However the concurrence is less than 1
ebit for the small amplitude, but it increases to one ebit for
larger amplitudes. Therefore, this state represents two classes of
entangled coherent states: the first is partial entangled states
and the second is maximum entangled  one (see \cite{allati1} for
more details).

\section{Entanglement through noise environment}

\subsection{Amplitude damping:Description}

\begin{figure}[b!]
\begin{center}
 \includegraphics[width=25pc,height=15pc]{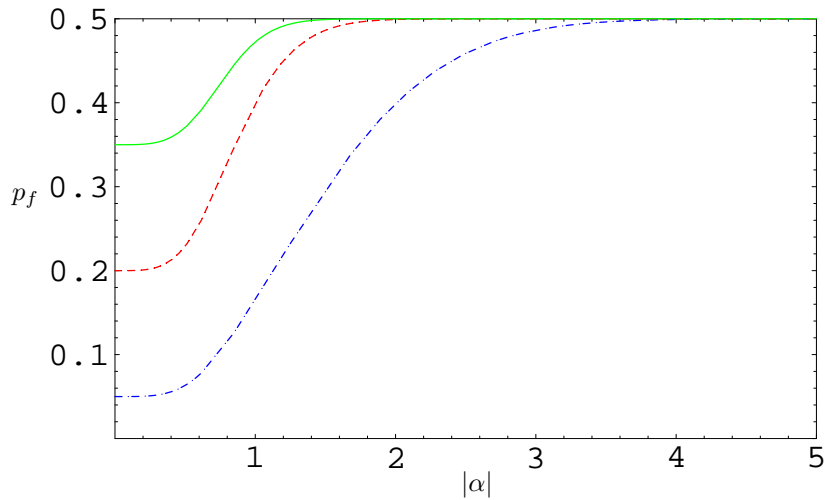}
 \put(-310,105){$p_{f}$}\put(-140,-5){$|\alpha|$}
  \caption{Phase flip probability $p_{f}$ as
   function of the field intensity $|\alpha|$. The dash-dot, dot and solid curves  for $\eta=0.9,0.6$ and
   $0.3$   respectively.}\label{phaseflip1}
\end{center}
\end{figure}
 In this section we  investigate the dynamics of the
maximum entangled state (\ref{max}), when it  passes through an
amplitude damping channel, which is defined  by a photon loss and
phase flip with probability $p_{f}$. The photon loss due to the
interaction of travelling  state (\ref{max})  with an optical
fiber prepared in a vacuum state. This interaction transfer the
state $\ket{\pm\alpha}\ket{0}_E$ to
$\ket{\pm\sqrt{\eta}\alpha}\ket{\pm\sqrt{1-\eta}\ket{\alpha}}_{E}$,
where $\eta$ is called the the transmissivity rate\cite{Ricardo}.
Tracing out the environment mode, one obtains a new state where
the amplitude is reduced from $\alpha$ to $\alpha\sqrt{\eta}$
\cite{Ricardo}. Therefore, the state vector
$\ket{\psi^{-}_{\alpha}}$ changes to $\ket{\psi^{-}_{\eta}}$
where,
\begin{equation}
\ket{\psi^{-}_{\eta}}=
\frac{1}{\sqrt{N_{\eta}}}(\ket{\sqrt{2}\alpha\sqrt{\eta},\alpha
\sqrt{\eta},\alpha \sqrt{\eta}}-\ket{-\sqrt{2}\alpha
\sqrt{\eta},-\alpha \sqrt{\eta},-\alpha \sqrt{\eta}}).
\end{equation}
On the other hand, if we assume that this travelling state is
subject to a phase noise with probability $p_f$, then the final
resulting effect is equivalent to the effect of the amplitude
damping channel. So, the final output state $\rho^{-}_{adc}$ which
is obtained  from the travelling state (\ref{max}) through
amplitude damping channel is given by,

\begin{equation}
\rho^{-}_{adc}=(1-P_{f})\rho^{-}_{\eta}+P_{f}Z\rho^{-}_{\eta}Z,
\end{equation}
where, $\rho^{-}_\eta=\ket{\psi^{-}_\eta}\bra{\psi^{-}_\eta}$  and
$Z$ is  the phase flip error, which is defined as $Z(\lambda_{1}|0
\rangle_{L}+\lambda_{2}|1\rangle_{L})=\lambda_{1}|0\rangle_{L}-\lambda_{2}|1\rangle_{L}$.
This operator  effects on the travelling state with probability
$p_f$ and with $(1-p_f)$ the state passes safely. In terms of
$\alpha$ and $\eta$, the probability $p_f$ is given by,
\begin{equation}
p_{f}=\frac{1-e^{-8|\alpha|^{2}}-e^{-4(1-\eta)|\alpha|^{2}}+e^{-4(1+\eta)|\alpha|^{2}}}{2(1-e^{-8|\alpha|^{2}})}.
\end{equation}
The behavior of the probability is shown in Fig.\ref{phaseflip1}
for different values of the transmissivity rate, $\eta$. It
displays that for small values of the field's intensity
$(\alpha\simeq 0)$, the minimum values of $p_f$ increases as the
noise strength $\eta$ decreases. In a small range of field
intensity $\alpha\in[0,4]$, $p_f$ increases faster and reaches its
maximum value ($\frac{1}{2})$ as the noise strength increases.
However, for larger values of the field intensity  the dynamics of
$p_f$ is independent of the noise strength, where
$p_f=\frac{1}{2}$. This means that for larger values of the
transmissivity rate $\eta\simeq 1$, the travelling state is almost
maximum and its resistance to  phase flip error is stronger.

\subsection{Dynamics of entanglement:three qubit}
To investigate the entanglement of a maximum entangled tripartite
state ( which is  defined by (\ref{max})), passes through
amplitude damping channel, we consider the following situation:
Let us assume that we have a source  supplies a three users,
Alice, Bob and Charlie with a maximum entangled state of  type
(\ref{max}). For simplicity, it is assumed that during the
transition from the source to the users, Bob and Charlie's qubit
are forced to pass through amplitude  damping  channel. According
to this suggested scenario, the dynamics of the travelling state
is given by,
\begin{equation}
\rho_{adc}=(1\otimes\mathcal{S}_1\otimes\mathcal{S}_2)\rho^{-}_{\alpha},
\end{equation}
where $\mathcal{S}_1$ and $\mathcal{S}_2$ represent the damping
channels which effect on Bob and Charlie's qubits respectively.
For simplicity we set $\mathcal{S}_1=\mathcal{S}_2=\mathcal{S}$
and rewrite the state vector $\ket{\psi^{-}_{\alpha}}$ by using
the orthogonal basis $u$ and $v$ defined as,
\begin{equation}\label{bas}
 \ket{\alpha}=\lambda_{\alpha}\ket{u}+\mu_{\alpha}\ket{v},\quad
\ket{-\alpha}=\lambda_{\alpha}\ket{u}-\mu_{\alpha}\ket{v},
\end{equation}
and
$\lambda_{\alpha}=(\frac{1+e^{-2|\alpha|^{2}}}{2})^{\frac{1}{2}}$
and $\mu_{\alpha}=(\frac{1-e^{-2|\alpha|^{2}}}{2})^{\frac{1}{2}}$.
Then the output state  vector can be written as,

\begin{eqnarray}\label{output}
|\psi_{out}\rangle&=&
(1+e^{i\theta})\Bigl\{\lambda_{\sqrt{2}\alpha}\lambda_{\alpha}^{2}\ket{uuu}+
\lambda_{\sqrt{2}\alpha}\lambda_{\alpha}\mu_{\alpha}\bigl(\ket{uuv}+\ket{uvu}\bigr)
+\lambda_{\sqrt{2}\alpha}\mu^2_{\alpha}\ket{uvv}
\nonumber\\
&+&
\mu_{\sqrt{2}\alpha}\lambda^{2}_{\alpha}\ket{vuu}+\mu_{\sqrt{2}\alpha}\lambda_{\alpha}\mu_{\alpha}\bigl(\ket{vuv}+\ket{vuu}\bigr)+\mu_{\sqrt{2}\alpha}\mu^{2}_{\alpha}\ket{vvv}\Bigr\}
\nonumber\\
&+&(1-e^{i\theta}\Bigl\{\lambda_{\sqrt{2}\alpha}\lambda_{\alpha}^{2}\ket{uuu}
-\lambda_{\sqrt{2}\alpha}\lambda_{\alpha}\mu_{\alpha}\bigl(\ket{uuv}+\ket{uvu}\bigr)
+\lambda_{\sqrt{2}\alpha}\mu^2_{\alpha}\ket{uvv}
\nonumber\\
&-&
\mu_{\sqrt{2}\alpha}\lambda^{2}_{\alpha}\ket{vuu}+\mu_{\sqrt{2}\alpha}\lambda_{\alpha}\mu_{\alpha}\bigl(\ket{vuv}+\ket{vuu}\bigr)
-\mu_{\sqrt{2}\alpha}\mu^{2}_{\alpha}\ket{vvv}\Bigr\}.
\end{eqnarray}
The lower bound of entanglement of  state $\rho_{out}$  can be
quantified by using a procedure described in \cite{Konrad}. This
procedure state that the concurrence for any two qubits state
$\ket{\zeta}\bra{\zeta}$ passes either in one or two sides of
channels $\mathcal{S}_1$ and $\mathcal{S}_2$ is bounded from above
in terms of the evolution of the concurrence of the maximally
entangled state under either one of the one-sided channels as:
\begin{equation}
\mathcal{C}\bigl[(\mathcal{S}_1\otimes\mathcal{S}_2)\ket{\zeta}\bra{\zeta}\bigr]=
\mathcal{C}\bigl[(\mathcal{S}_1\otimes\mathcal{S}_2)\ket{\phi}\bra{\phi}\bigr]\mathcal{C}\bigr[\ket{\zeta}\bra{\zeta}\bigl],
\end{equation}
where $\ket{\phi}\bra{\phi}$ is  a maximum entangled  two qubits
state. For a three qubits state we use the same procedure, where
we consider  GHZ  state represent the maximum entangled state.
Therefore the concurrence of  the maximum entangled state
(\ref{max}) is bounded from the above as \cite{Simon},
\begin{equation}\label{K}
\mathcal{C}^{23/1}[(1\otimes S\otimes S)\rho^{-}_{\alpha}]\leq
\mathcal{C}^{23/1}[(1\otimes S\otimes S)|GHZ\rangle\langle
GHZ|]\mathcal{C}^{23/1}[\rho^{-}_{\alpha}].
\end{equation}
To quantify the degree of entanglement of the output state
$\rho_{out}=\ket{\psi_{out}}\bra{\psi_{out}}$, we have to
reexpress  the GHZ in the new basis $u$ and $v$ as,

\begin{equation}
\ket{GHZ}=\frac{1}{\sqrt{2}}(\ket{uuu}+\bra{vvv}).
\end{equation}
As a first step, we consider one side effect of the amplitude
damping channel on the GHZ state. This evolution is defined as,
\begin{equation}
(1\otimes 1\otimes S)\ket{GHZ_{uv}}\bra{GHZ_{uv}}=\left(%
\begin{array}{cccccccc}
  a & 0 & 0 & 0 & 0 & 0 & 0 & f \\
  0 & b & 0 & 0 & 0 & 0 & e & 0 \\
  0 & 0 & 0 & 0 & 0 & 0 & 0 & 0 \\
  0 & 0 & 0 & 0 & 0 & 0 & 0 & 0 \\
  0 & 0 & 0 & 0 & 0 & 0 & 0 & 0 \\
  0 & 0 & 0 & 0 & 0 & 0 & 0 & 0 \\
  0 & e^{*}& 0 & 0 & 0 & 0 & c & 0 \\
  f^{*}& 0 & 0 & 0 & 0 & 0 & 0 & d ,\\
\end{array}
\right),
\end{equation}
where,
\begin{eqnarray}
a&=&P_{s}\frac{\lambda^{2}_{\sqrt{\eta}u}}{4\lambda^{2}_u}, \quad
b=(1-P_{s})\frac{\mu^{2}_{\sqrt{\eta}u}}{4\mu^{2}_u}, \quad
f=P_{s}\frac{\mu_{\sqrt{\eta}u}\nu_{\sqrt{\eta}u}}{4\mu_{u}\nu_{u}},
\nonumber\\
e&=&-(1-P_{s})\frac{\lambda_{\sqrt{\eta}u}\mu_{\sqrt{\eta}u}}{4\lambda_{u}\mu_\alpha},\quad
c=(1-P_{s})\frac{\lambda^{2}_{\sqrt{\eta}\alpha}}{4\lambda^{2}_{u}},
\quad d=P_{s}\frac{\mu^{2}_{\sqrt{\eta}u}}{4\lambda^{2}_u}.
\nonumber\\
P_{s}&=&\frac{1}{2}+\frac{e^{-4(1-\eta)|u|^{2}}-e^{-4(1+\eta)|u|^{2}}}{2(1-e^{-8|u|^{2}})}.
\end{eqnarray}
It is clear that, the outer and the inner elements of the state
represent the  "unflipped" and "flipped" GHZ states of reduced,
$\sqrt{\eta}u$ amplitude respectively. Then the dynamics of GHZ
state through two-sides amplitude damping channel is given by,
\begin{eqnarray}
(1\otimes S\otimes
S)\ket{GHZ_{u,u,u}}\bra{GHZ_{u,u,u}}&=&P_{s}\ket{GHZ_{u,\sqrt{\eta}u,\sqrt{\eta}u}}\bra{
GHZ_{u,\sqrt{\eta}u,\sqrt{\eta}u}}
\nonumber\\
&+&(1-P_{s})Z\ket{GHZ_{u,\sqrt{\eta}u,\sqrt{\eta}u}}\bra{GHZ_{u,\sqrt{\eta}u,\sqrt{\eta}u}}Z.
\end{eqnarray}
\begin{figure}
\begin{center}
  \includegraphics[width=25pc,height=15pc]{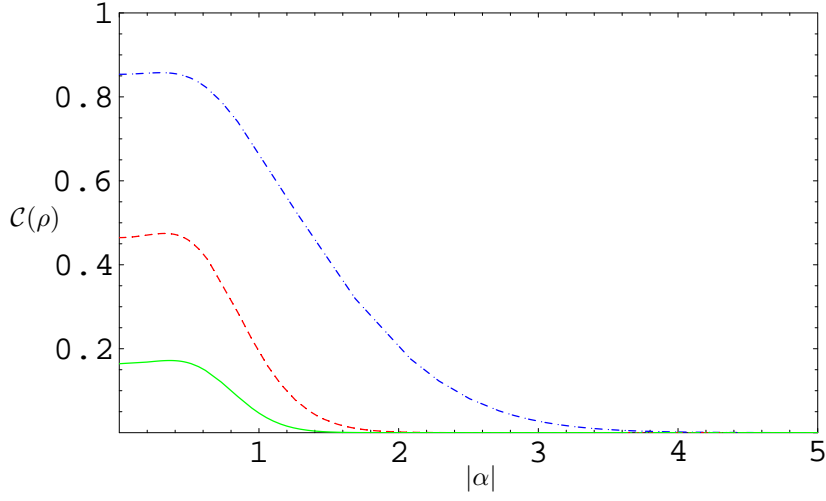}
  \put(-3105,105){$\mathcal{C}(\rho)$}\put(-140,-5){$|\alpha|$}
  \put(-312,93){$\mathcal{C}(\rho)$}
  \caption{The concurrence $\mathcal{C}(\rho)$ as a function of $\eta$ and $|\alpha|$. The dash-dot, dot and solid curves are
  evaluated at $\eta=0.9,0.6$ and $0.3$ respectively.}\label{Con}
\end{center}
\end{figure}
The  concurrence of the travelling state (\ref{output}) through
the amplitude damping channel is given by,
\begin{equation}\label{conc}
\mathcal{C}(\rho)=2max[0,|e|-\sqrt{ad},|f|-\sqrt{bc}].
\end{equation}
Fig.\ref{Con} shows the dynamics of the concurrence
$\mathcal{C}(\rho)$ for different values of the the transmissivity
rate $\eta$. If the travelling state through the amplitude damping
channel is partially entangled state i.e. $\eta$ is small, the
entanglement, which is represented by the concurrence, is very
small and vanishes for small values of the field intensity.
However, for larger values of $\eta$, the initial entanglement is
large and decreases smoothly  as  the field's intensity increases.
So, to keep the entanglement of the MMECS  over the amplitude
damping channel survival for a long time, one has to decrease the
field's intensity. It is clear that, for larger values of the
transmissivity rate $\eta$ the travelling state is more robust.

\subsection{Dynamics of entanglement:"$m$ modes}
In this section, we assume that the users share a coherent state
of $m$ modes given by,

\begin{eqnarray}  \label{gen}
\ket{\Psi^{\pm}_{0...m}} &=& A^{\pm}_{m+1}\Bigl(|2^{\frac{m-1}{2}%
}\alpha\rangle_{0}...|2^{\frac{1}{2}}\alpha\rangle_{m-2}|\alpha%
\rangle_{m-1}|\alpha\rangle_{m}  \nonumber \\
&&\pm|-2^{\frac{m-1}{2}}\alpha\rangle_{0}...|-2^{\frac{1}{2}%
}\alpha\rangle_{m-2}|-\alpha\rangle_{m-1}|-\alpha\rangle_{m}|\Bigr),
\end{eqnarray}
where $A^{\pm}_{m+1}=[2(1\pm
e^{-2^{m+1}|\alpha|^{2}})]^{-\frac{1}{2}}$, is the normalized
factor. This state can be generated from Schr\"{o}dinger state and
optics devices. In \cite{allati1}, we have shown that this state
represents a quantum network, shared between multiusers where  one
user called  emitter posses the mode $0$ and the other users share
the remaining $m$ modes. Moreover we have employed this state to
teleport a multipartite states of $m$ modes. The degree of
entanglement of the  network which is defined by the state
$\rho_{gen}=\ket{ \psi^{\pm}_{0,...m}} \bra{\psi^{\pm}_{0,...,m}}$
is given by  \cite{allati2},
\begin{equation}
C^{0/1,2,...,m}=1.
\end{equation}
for $\theta=\pi$ or for $\theta=0$.

The main aim of this section is investigating the entangled and
separable properties of this multipartite state. Let us assume
that there are $m$ modes of the state (\ref{gen}) passes through
an  amplitude damping channel. In this case the output state can
be written as,
\begin{equation}
\rho^{\pm}_{out}=(1-p_{f,m})\rho^{\pm}_{\eta,0...m}+p_{f,m}Z\rho^{\pm}_{\eta,0,...,m}Z,
\end{equation}
where,
$\rho^{\pm}_{out}=\ket{\Psi^{\pm}_{0,...,m}}\bra{\Psi^{\pm}_{0,...,m}}$
and $p_{f,m}$ is the probability that the phase flip affects the
travelling state through the amplitude damping channel. This
probability is given by,
\begin{equation}\label{prob}
p_{f,m}=\frac{1-e^{-2^{m}|\alpha|^{2}}-e^{-2^{m-1}(1-\eta)|\alpha|^{2}}
+e^{-2^{m-1}(1+\eta)|\alpha|^{2}}}{2(1-e^{-2^{m}|\alpha|^{2}})},
\quad m\geq1
\end{equation}
\begin{figure}
  \begin{center}
 \includegraphics[width=16pc,height=10pc]{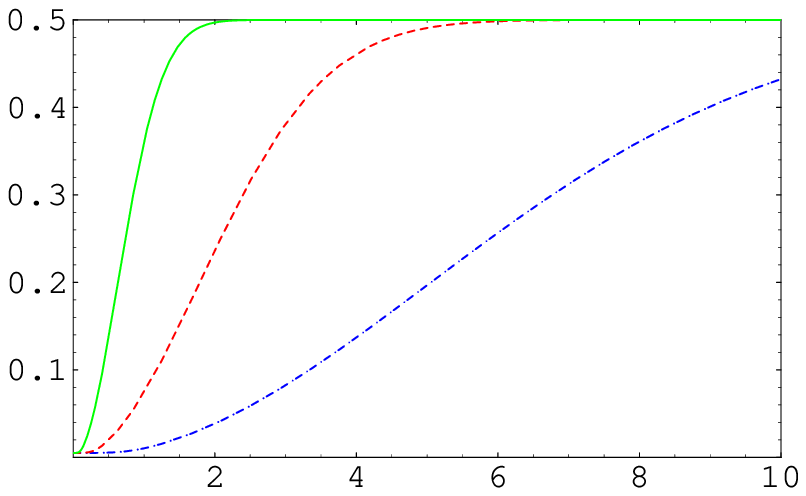}~\quad
  \includegraphics[width=16pc,height=10pc]{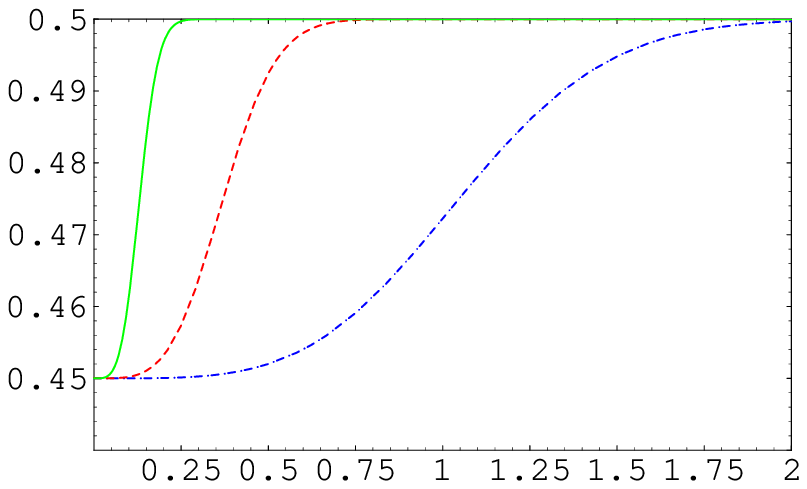}
\put(-415,70){$p_{f,m}$} \put(-200,70){$p_{f,m}$}
\put(-300,-10){$|\alpha|$} \put(-90,-10){$|\alpha|$}
\put(-230,105){$(a)$}\put(-20,100){$(b)$}
    \caption{The probability of the phase flip error $p_{f,m}$
  which effects on the travelling state (\ref{gen}). The dash-dot,
  dot and solid curves for $m=2,5,8$ respectively and the transmissivity rate(a)
  $\eta=0.99$ and (b)  $\eta=0.1$.}\label{PFN}
  \end{center}
\end{figure}
Fig.(\ref{PFN}a),  displays the  dynamics of the probability
${p_f}$ of the phase bit flip error which effects on the
travelling state through the amplitude damping channel for
different values of the modes while transmissivity rate is large
$(\eta=0.99)$, i.e. the travelling state is almost maximum. It is
clear that, for small vales of modes, the probability $p_{f,m}$
increases gradually to reach its maximum value $(=0.5)$ for lager
values of the field intensity $|\alpha|$. However for larger
values of $m$, $p_{f,m}$ increases abruptly and reaches the
maximum bound for small values of the field intensity. In
Fig.(\ref{PFN}b), we assume that the travelling state (\ref{gen})
through the amplitude damping channel is partially entangled
state, where we set  the transmissivity rate $\eta=0.1$. In this
case, the resistance of the input state (\ref{gen}) for the phase
bit flip error is very fragile, where $p_{f,n}$ reaches its
maximum values for smaller values of the field intensity.

To quantify the degree of entanglement contained in the travelling
state (\ref{gen}) through the amplitude damping channel, we
rewrite the lower bound of entanglement to include $m$ modes.
Therefore Eq.(\ref{K}) can be generalized as,

\begin{equation}\label{c-gen}
C[(1\otimes ...\otimes S\otimes S)\rho^{\pm}_{gen}]\leq
C[(1\otimes...\otimes S\otimes
S)|GHZ_{\alpha,...,\alpha}\rangle\langle
GHZ_{\alpha,...,\alpha}|]C^{1/2...m}[\rho^{\pm}_{gen}].
\end{equation}
To evaluate this bound of entanglement, one has to investigate the
effect of the amplitude noise channel on the  $m+1$ modes of GHZ
state which in  the orthogonal basis takes the form,
\begin{equation}
\ket{GHZ}=\frac{1}{\sqrt{2}}(\ket{u...u}+\ket{v...v}).
\end{equation}
The dynamics of  $\ket{GHZ}$  state through the amplitude damping
channel is given by,
\begin{eqnarray}
(1\otimes...\otimes S\otimes
S)\ket{GHZ_{\alpha,\alpha,...,\alpha}}\bra{
GHZ_{\alpha,\alpha,...,\alpha}}&=&(1-P_{f,m})\ket{GHZ_{\alpha,\sqrt{\eta}\alpha,...,\sqrt{\eta}\alpha}}\bra{
GHZ_{\alpha,\sqrt{\eta}\alpha,...,\sqrt{\eta}\alpha}}\nonumber\\
&&+P_{f,m}Z\ket{GHZ_{\alpha,\sqrt{\eta}\alpha,...,\sqrt{\eta}\alpha}}\bra{
GHZ_{\alpha,\sqrt{\eta}\alpha,....,\sqrt{\eta}\alpha}}Z.
\nonumber\\
\end{eqnarray}
The amount of entanglement is quantified by means of the
concurrence as,

\begin{equation}
\mathcal{C}_{\pm}=\frac{1-2\mathcal{P}_{f,m}}{1\pm
exp\Bigl\{-2^{m-1}(1+\eta)|\alpha|^2\Bigr\}}\sqrt{1-exp(-2^m
|\alpha|^2)}\sqrt{1-exp(-2^m\eta|\alpha|^2)},
\end{equation}
where $\mathcal{C}_{+}$  and $\mathcal{C}_{-}$ for $\theta=0,\pi$
respectively and the concurrence
$\mathcal{C}^{1/2,...m}[\rho^{\pm}_{gen}]=1$(see Eq.(19)).
\begin{figure}[t!]
  \begin{center}
 \includegraphics[width=16pc,height=10pc]{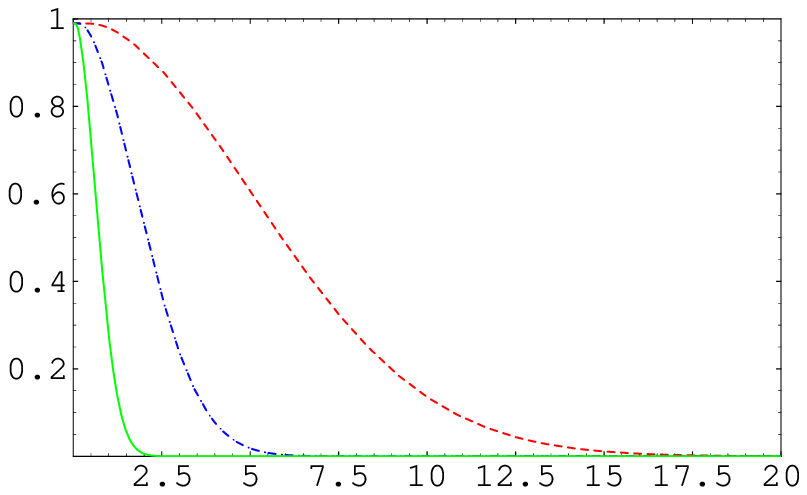}~\quad
  \includegraphics[width=16pc,height=10pc]{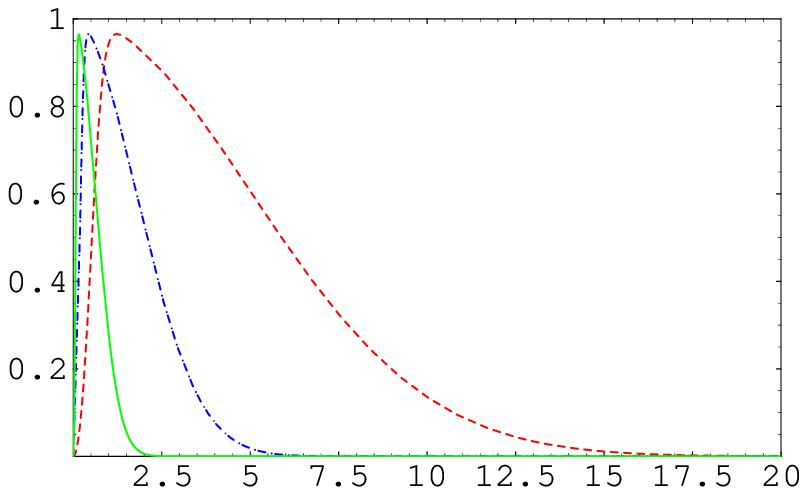}
\put(-415,65){$\mathcal{C}_{+}$} \put(-200,65){$\mathcal{C}_{-}$}
 \put(-230,100){$(a)$}\put(-20,100){$(b)$}
 \put(-300,-8){$|\alpha|$}
\put(-90,-8){$|\alpha|$}
    \caption{Dynamics of the concurrence for different values of modes. The dot, dash-dot and solid curves for
    $m=2,5$, $8$ respectively and transmissivity rate $\eta=0.9$ (a) For $\theta=\pi$(b) For $\theta=0$.}\label{ConcG99}
  \end{center}
\end{figure}

The dynamics of  entanglement which is represented by concurrence
for different values of the phase $\theta$ is described in
Fig.\ref{ConcG99}, where the transmissivity rate $\eta$ is assumed
to be fixed. It is clear that, for odd state i.e. $\theta=\pi$,
the concurrence decreases as the field intensity increases as
shown in Fig.(\ref{ConcG99}a). The decay of entanglement depends
on the number of modes of the travelling state. For small values
of modes, the entanglement decays smoothly and gradually to
vanishes completely at larger values of the field intensity.
However as the number of modes increases the entanglement decays
fast and abruptly  vanishes at  small values of the field
intensity. The dynamics of the concurrence for an even class of
MMECS, is shown in Fig.(\ref{ConcG99}b), where $\theta=0$. It is
clear that for $|\alpha|=0$, the travelling state is almost
separable. However as soon as $|\alpha|$ increases, the
entanglement increases sharply to reach its maximum value in a
very small range of the $|\alpha|$ depending on the number of
travelling modes. However, as $|\alpha|$ increases more, the
entanglement decays gradually for small values of $m$ and hastily
for larger values of $m$. From Figs.(\ref{ConcG99}$a
\&\ref{ConcG99}b$),the entanglement vanishes for the same value of
$|\alpha|$. Therefore, the amount of entanglement contained in the
odd and even travelling states over the amplitude damping channel
vanishes  for the same value of the field intensity.

\begin{figure}
  \begin{center}
 \includegraphics[width=16pc,height=10pc]{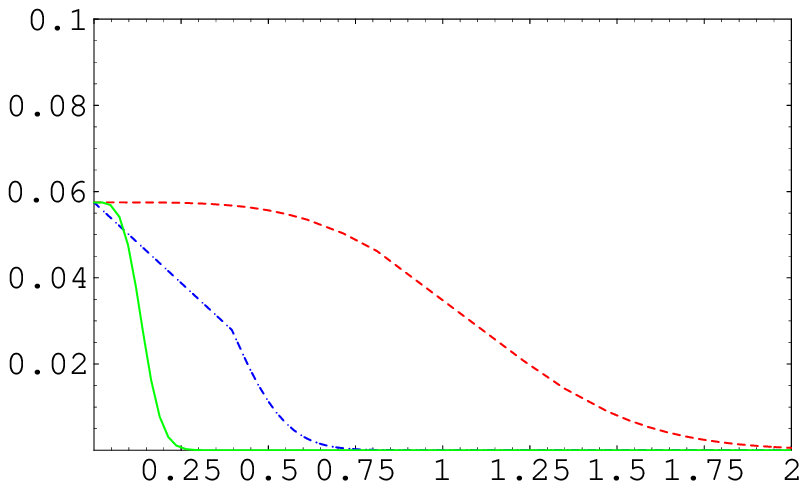}~\quad
     \includegraphics[width=16pc,height=10pc]{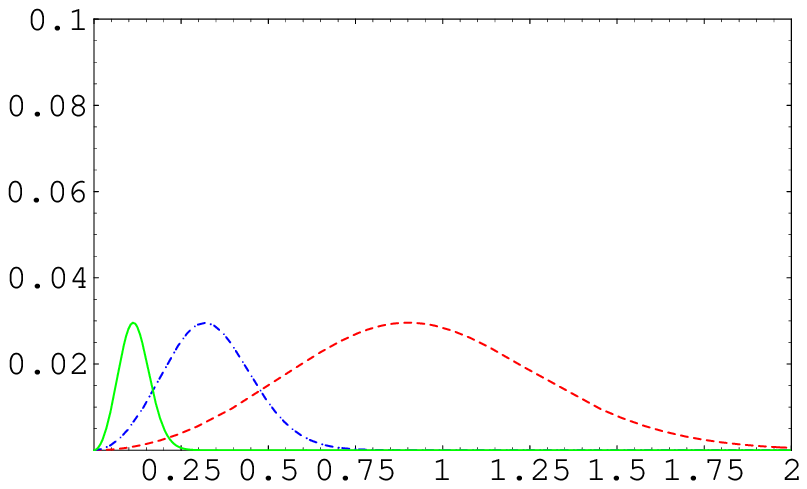}
  \put(-230,100){$(a)$}\put(-20,100){$(b)$}
\put(-415,65){$\mathcal{C}_{+}$} \put(-200,65){$\mathcal{C}_{-}$}
\put(-300,-8){$|\alpha|$} \put(-90,-8){$|\alpha|$}
    \caption{The same as Fig.\ref{ConcG99}, but
    $\eta=0.1$.}\label{ConcG1}
  \end{center}
\end{figure}

Fig.(\ref{ConcG1}) shows the dynamics of entanglement for small
value of $\eta(=0.1)$, i.e the travelling state over the amplitude
damping channel has an initial small value of entanglement. The
general behavior is the same as that depicted in
Fig.(\ref{ConcG99}). However, the  initial amount of entanglement
is very small and vanishes very  fast at small values of the field
intensity.

\section{Conclusion}

The dynamics of a  maximum entangled state passes through an
amplitude damping channel is discussed. We showed that,the
entanglement decays gradually  for larger values of the field
intensity and small values of the transmissivity rate. However For
small values of the transmissivity rate, the entanglement vanishes
at small  values of the field intensity. Therefore to increase the
resistance of the MMECS to entanglement degradation one has to
increase the field's intensity when the transmissivity rate is
large.

The dynamics of a multi-modes  entangled state passes through an
amplitude damping channel is investigated. This type of study
displays the effect of the noise strength, the phase flip operator
and the field intensity. We show that the travelling state
suffering from the phase flip effect with high probability for
small values   the noise strength absorption parameter. However
the robustness of this multi-modes entangled state for the phase
flip operator, decreases as  the photon absorption decreases,
where in this case the travelling state is partially entangled
state. Moreover, this resistance  decreases as the field's
intensity increases. On the other hand, the probability of the
phase error effects depends on the number of photons for each
mode, where the probability is maximized as the number of photons
increases.

 The entanglement of MMECS  for different  modes is investigated,
 where the entanglement decreases gradually for small values of
 modes. However as the number of modes  increases, the
 entanglement decays very fast for small values of the field's
 intensity. It is shown that the entanglement for  both the  odd and even MMECS states
 completely vanishes at the same values of the field intensity.
 The decay rat of the travelling entanglement depends on the
 field's  intensity and the transmissivity rate.

\textbf{Acknowledgement}: we are grateful for the helpful comments
given by the referees which improves our results.

\end{document}